\documentclass[final,3p,times,twocolumn]{elsarticle}
\usepackage{graphicx}
\usepackage{amsmath}
\usepackage{amssymb}
\usepackage{hyperref}
\usepackage{color}

\begin{document}

\title{Interface induced exchange bias effect in La$_{0.67}$Sr$_{0.33}$MnO$_3$/SrIrO$_3$ multilayer}

\author{Rachna Chaurasia}
\author{K. C. Kharkwal}
\author{A. K. Pramanik\corref{cor1}}
\ead{akpramanik@mail.jnu.ac.in}

\address{School of Physical Sciences, Jawaharlal Nehru University, New Delhi - 110067, India.}

\begin{abstract}
Epitaxial superlattices of ferromagnetic/paramagnetic La$_{0.67}$Sr$_{0.33}$MnO$_3$/SrIrO$_3$ materials have been prepared on SrTiO$_3$ (100) substrate using pulse laser deposition technique. An unexpected onset of interface magnetic interaction has been observed around 40 K. Interestingly, magnetic exchange bias effect has been observed in both field cooled and zero field cooled magnetization loops, however, the shifting of loop is opposite in both measurements. Exchange bias field vanishes as temperature increases to interface magnetic ordering temperature. Moreover, exchange bias field is found to decrease with increasing cooling field. We believe that tuning of magnetic exchange at interface during field cooling induces this evolution in nature of exchange bias field.
\end{abstract} 


\maketitle
\section {Introduction}
In recent years, physics related to spin-orbit coupling (SOC) effect has drawn much attention in condensed matter physics. Normally, the electron correlation ($U$) and SOC exhibits a comparable energy scale in heavy transition metal oxides (TMOs), and Ir based oxides are the best example for that.\cite{cao1,rau,krempa} In parallel to bulk materials, focus has also been paid on thin films and heterostructures as 2 dimensional (D) confinement of electrons drastically change the physical properties. In addition, interfacial strain, oxygen vacancies and synthesis parameters play crucial role toward its modified physical properties. \cite{biswas,hao,lee,lupascu,jenderka,liu,schutz,anderson} Thin films are not only useful for their various applications such as, information storage, spintronics, etc. also they are well known for many interesting phenomena such as unusual magnetic behavior, superconductivity, ferroelectricity, etc.

Perovskite based SrIrO$_3$ is of particular interest. Bulk SrIrO$_3$ shows paramagnetic (PM) and metallic (M) behavior throughout the temperature ($T$) range.\cite{blanchard} However, this material lies at the verge of ferromagnetic (FM) instability which has been indicated from band structure calculations\cite{zeb} as well as from chemical doping that shows even small amount of chemical substitution induces ferromagnetism with high ordering temperature ($T_c$).\cite{cui,qasim} There are several reports for SrIrO$_3$ thin films. The studies show films of SrIrO$_3$ evolve from metallic to insulating both with decreasing film thickness as well as with varying strain realized from different substrates.\cite{biswas,groenendijk} The angle resolved photoemission spectroscopy (ARPES) studies on SrIrO$_3$ films have illustrated an interesting aspect of band structure which evolves with both octahedral rotation and dimensionality of material.\cite{liu,schutz} Recently, few attempts are made to fabricate heterostructures of SrIrO$_3$ with different magnetic materials. A recent study has shown an interface induced perpendicular magnetic anisotropy in (La$_{1-x}$Sr$_x$MnO$_3$)/(SrIrO$_3$) superlattices which tunes with rotation of oxygen octadedra at interface.\cite{yi} Similarly, an interface driven topological Hall effect, which can be controlled through interface Dzyaloshinskii-Moriya interaction, has been observed in SrRuO$_3$-SrIrO$_3$ bilayer.\cite{matsuno}

Here, we have studied an interface induced modification of magnetic anisotropy through exchange bias (EB) effect in superlattices made of 3$d$ and 5$d$ oxides i.e., La$_{0.67}$Sr$_{0.33}$MnO$_3$ (LSMO) and SrIrO$_3$ (SIO) which are FM and PM, respectively. EB is normally manifested through shifting of magnetic hysteresis loop in a system with an interface between different magnetic states when cooled under magnetic field ($H$).\cite{nogues} An unidirectional magnetic anisotropy is induced at interface due to field cooling process. Conventionally, hysteresis loop shifts toward negative field direction when cooled in positive field which is called as negative EB effect and is observed in most of the systems. However, there are few examples where an asymmetry in $M(H)$ loop has been observed toward positive field axis even after cooling in zero magnetic field, which is rather surprising and unconventional.\cite{shang,zhang,nogues1} Since its discovery, EB effect has been studied extensively following its possible application in spintronics. Obviously, artificially fabricated superlattices offer an ideal system to study the EB effect not only because an individual layer of particular magnetic state can be chosen but also interface plays a crucial role here with its easy modified magnetic character. Interface induced EB behavior have been reported in many studies in 3$d$-4$d$ systems\cite{solignac,sow,he,ke} but it remains mostly unexplored for 3$d$-5$d$ based oxide superlattices.

Our results indicate an onset of interface magnetic interaction below around 40 K. Field dependent magnetization data $M(H)$ measured under both field cooled (FC) and zero field cooled (ZFC) condition show an asymmetric behavior which suggests an EB behavior. The shifting of $M(H)$ loop is, however, opposite in these both measurements which implies both negative and positive EB effects can be present in this superlattices. The EB effect vanishes when temperature is raised to interface magnetic ordering temperature. Similarly, EB field decreases and coercive field increases with cooling field.   

\section {Experimental Details}
Two epitaxial heterostructures of [SIO/LSMO]$_n$ with $n$ = 3 and 6 are grown on single crystal SrTiO$_3$ (100) substrate using pulse laser deposition (PLD) technique with KrF laser (248 nm). Respective target materials of LSMO and SIO have been prepared using conventional solid state reaction method. Deposition of both the materials are done at 750$^o$C with laser pulse repetition rate 5 Hz and energy and 1.2 J/cm$^2$. During deposition, a constant flow of oxygen is maintained at 0.1 mbar. First, a layer of LSMO has been deposited and on its top SIO is deposited. This bi-layer is repeated three times with individual layer thickness 20 nm (labelled as ML$_{[20/20]3}$) and six times with an individual layer thickness around 15 nm (labelled as ML$_{[15/15]6}$). The thickness of LSMO is safely chosen so that it can support fully developed FM state.\cite{huijben} After deposition, the chamber has been filled with oxygen at pressure $\sim$ 500 mbar and cooled to room temperature. Epitaxial quality of the film is checked using x-ray diffraction (XRD) with CuK$_\alpha$ radiation (PANalytical). All the magnetic measurements on sample ML$_{[20/20]3}$ are performed using SQUID (Quantum Design) and FC M(H) of sample ML$_{[15/15]6}$ and LSMO thin film has been recorded using VSM (Cryogenic Limited).

\begin{figure}
	\centering
		\includegraphics[width=8cm]{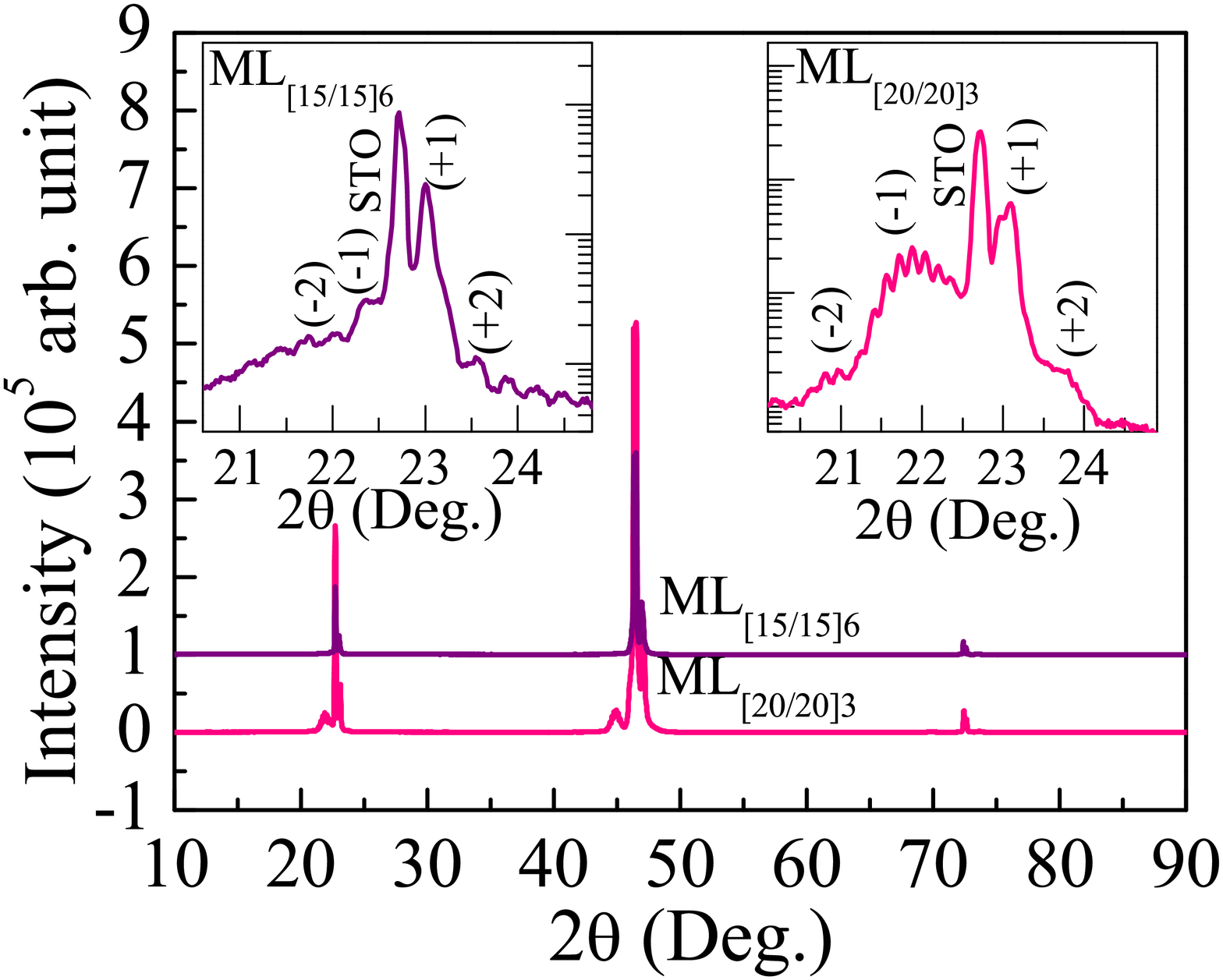}
	\caption{(color online) X-ray diffraction pattern of ML$_{[20/20]3}$ and ML$_{[15/15]6}$ multilayer grown on STO (100). Magnified view of (100) x-ray diffraction peak has been shown for ML$_{[15/15]6}$ and ML$_{[20/20]3}$ multilayer in left and right inset, respectively. positive and negative index marked peaks represents LSMO and SIO, respectively.}
	\label{fig:Fig1}
\end{figure}

\section{Results and Discussions}
Figure 1 shows typical $\theta$-2$\theta$ XRD scan of [LSMO/SIO]$_3$ and [LSMO/SIO]$_6$ multilayer. It is clear in figure that only (100) reflections related to substrate are observed, however, satellite peaks due to films near the respective substrate peaks are also present. Left and Right inset of Figure 1 shows expanded view of low angle XRD data for ML$_{[15/15]6}$ and ML$_{[20/20]3}$ multilayer, respectively showing diffraction peaks related to substrate SrTiO$_3$ as well as deposited materials i.e., LSMO and SIO. The SIO and LSMO peaks are observed at left and right side of the substrate peak, and marked with positive and negative index respectively. The figure shows Bragg's reflection with two order of satellite peaks and well resolved thickness fringes related to SIO and LSMO peaks. This indicates that present multilayers is of good quality and has an epitaxial structure with substrate. We calculate psuedocubic lattice parameter of STO and LSMO is 3.90 and 3.88 {\AA}, respectively which indicates $\sim$ 0.5\% of lattice mismatch between substrate and first layer of LSMO. The total thickness $D$ of the multilayer has been calculated by the following formula,\cite{kishore}

\begin{eqnarray}
 D = \frac{(m-n)\lambda}{2(sin\theta_m-sin\theta_n)}
\end{eqnarray}

where $\theta_m$ and $\theta_n$ are position of $m$ and $n$ order peaks and $\lambda$ is the wavelength of x-ray used. Using Eq. 1, we have calculated the thickness about 125 and 174 nm for ML$_{[20/20]3}$ and ML$_{[15/15]6}$ multilayer, respectively. These calculated thickness for both multilayers are close to the values that are expected from growth rate estimation.\cite{kishore}

\begin{figure}
	\centering
		\includegraphics[width=8cm]{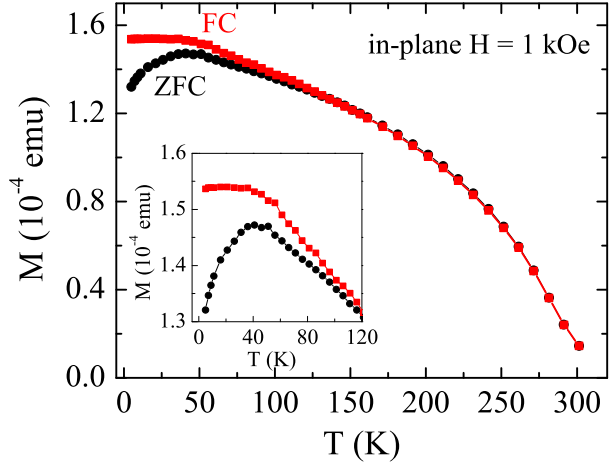}
	\caption{(color online) Magnetization data measured under ZFC and FC protocol in 1 kOe field are shown as a function of temperature for [LSMO/SIO]$_3$ (ML$_{[20/20]3}$) multilayer.}
	\label{fig:Fig2}
\end{figure} 

Figure 2 shows $M(T)$ data of present heterostructure measured in 1 kOe field applied parallel to the plane of the film, following ZFC and FC protocol. With lowering the temperature, moment is seen to increase in both ZFC and FC data which is due to LSMO component as the $T_c$ of bulk LSMO is above room temperature.\cite{urushibara} At low temperature, however, $M_{ZFC}$ decreases showing a peak around 40 K ($T_P$) while $M_{FC}$ apparently saturates below $T_P$. Data further show a large bifurcation between ZFC and FC magnetization which is prominent below $T_P$ though starts at $T > T_P$. While bulk LSMO and SIO are respectively FM and PM, this downfall in $M_{ZFC}$ at low temperatures is quite interesting. This behavior in $M(T)$ is likely caused by an interface effect. The interface where LSMO and SIO meet is magnetically modified and the two transition metals Mn and Ir engage in AFM type interaction via oxygen. We believed that while though bulk SIO is PM over the temperature, it develops FM ordering at low temperature which facilitates an AFM type spin coupling at interface. The low temperature FM ordering in SIO is, however, not caused by a proximity effect due to neighbouring LSMO layer. The SIO lies at the verge of FM instability where the previous studies show a development of FM behavior with a small amount of chemical substitution in SIO.\cite{zeb,cui,qasim} In present multilayer, we believe that strain arising from substrate and interface induces FM behavior in SIO at low temperature. The similar behavior has previously been observed at La$_{0.67}$Sr$_{0.33}$MnO$_3$/SrRuO$_3$ (FM/FM) interface.\cite{solignac,he,ke}

Figure 3(a) exhibits $M(H)$ data of ML$_{[20/20]3}$ heterostructure measured at 5 K in field range of $\pm$ 10 kOe (in-plane magnetic field ($H$)) under ZFC and FC protocol. In ZFC measurement, the sample has been cooled in zero field to a specific temperature and $M(H)$ is measured. For FC measurement, the sample is cooled in cooling field ($H_{cool}$) to a specific temperature and the field has been scanned from +$H_{cool}$ to -$H_{cool}$ and returned to +$H_{cool}$. As evident in Figure 3(a), both ZFC and FC $M(H)$ data show almost saturation and loop closing which suggest both the $M(H)$ loops are not minor hysteresis. To further check the saturation of FC and ZFC $M(H)$ loops, we have used a criterion\cite{harres} where first and second derivatives of $M(H)$ plot (i.e., d$M$/d$H$ and d$^2$$M$/d$H$$^2$, respectively) have been evaluated as shown in Figure 3(b) and 3(c), respectively. The ascending and descending branches of both derivatives coincide in high field regime, which conclusively shows that collected $M(H)$ loops are not minor hystersis. However, the moment in FC $M(H)$ shows a comparatively higher value which is possibly due to AFM type coupling at interface. The upper inset of Figure 3 shows an expanded view of $M(H)$ close to origin which clearly indicates an asymmetric type hysteresis loop for both ZFC and FC $M(H)$ data. For instance, in case of ZFC data the $M(H)$ is both shifted horizontally toward positive field axis and vertically toward negative moment axis. We obtain corresponding coercive field values $H_c^L$ = -115 and $H_c^R$ = 241 Oe and remnant magnetization values $M_r^U$ = 1.193 $\times$ 10$^{-4}$ and $M_r^D$ = 1.318 $\times$ 10$^{-4}$ emu. For FC data, we again observe both horizontal and vertical shift of $M(H)$ plot, however, the direction of shift is opposite to ZFC data yielding values $H_c^L$ = -289, $H_c^R$ = 161 Oe and $M_r^U$ = 1.754 $\times$ 10$^{-4}$, $M_r^D$ = 1.320 $\times$ 10$^{-4}$ emu. Here, $H_c^L$ and $H_c^R$ are the negative and positive field values where $M$ = 0, and $M_r^U$ and $M_r^D$ are the upper and down values of remnant magnetization at $H$ = 0. This asymmetric behavior of FM hysteresis in Figure 3 is not due to individual component (LSMO or SIO) as pure FM materials contribute only to symmetric $M(H)$ loop. The asymmetry type of $M(H)$ loop is commonly seen in case of exchange bias (EB) effect. The nature shifting as well as closing of $M(H)$ in Figure 3 suggest this asymmetry is caused by an EB effect.

\begin{figure}
	\centering
		\includegraphics[width=8cm]{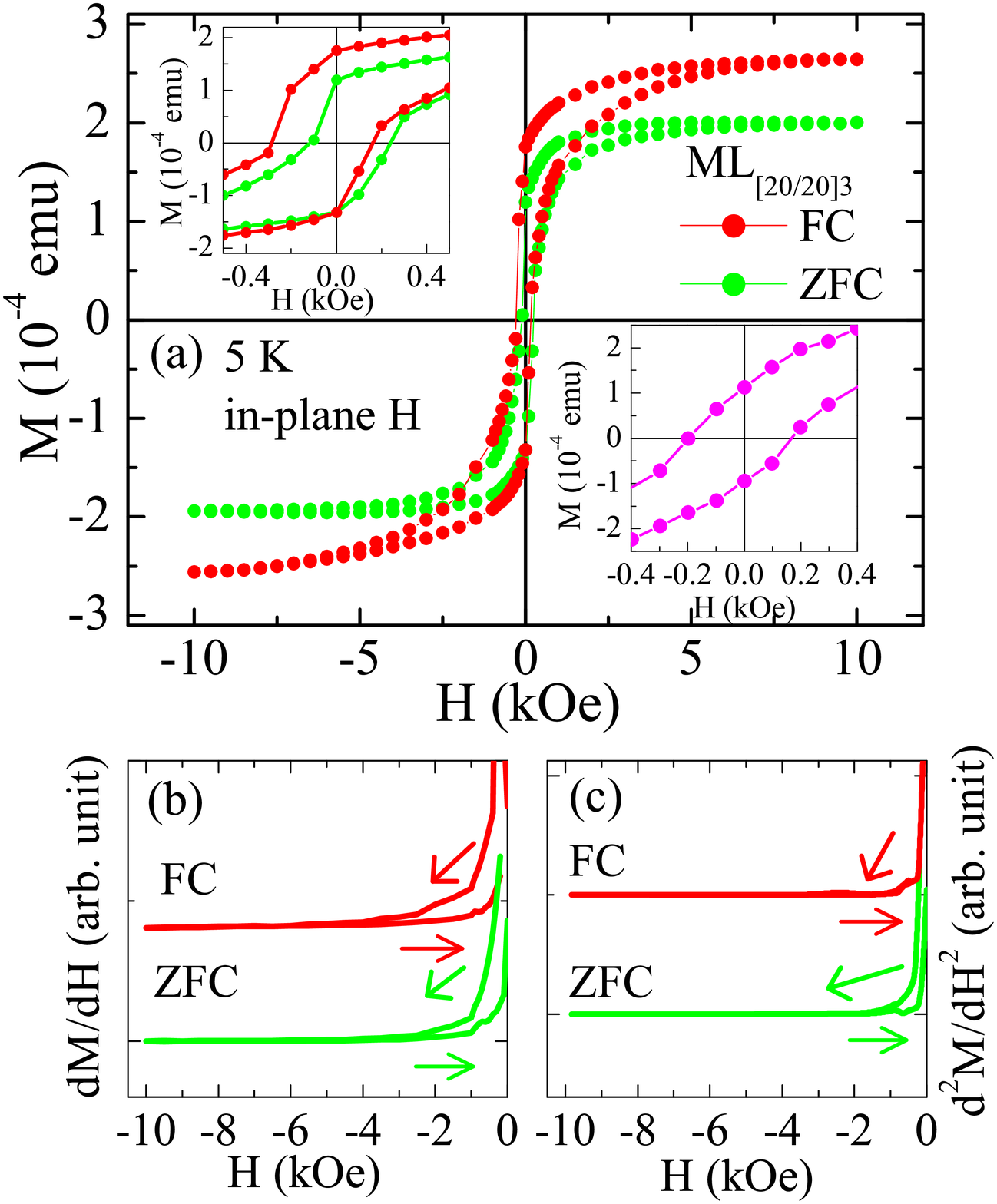}
	\caption{(color online) (a) Field dependent magnetization measured at 5 K under ZFC and FC protocol are shown for ML$_{[20/20]3}$ multilayer. Upper inset shows an expanded view close to origin. Lower inset shows an expanded view of FC $M(H)$ data for LSMO films at 5 K. (b) First derivative  and (c) second derivative of ascending and descending branches of FC and ZFC $M(H)$ data.}
	\label{fig:Fig3}
\end{figure}

\begin{figure}
	\centering
		\includegraphics[width=8cm]{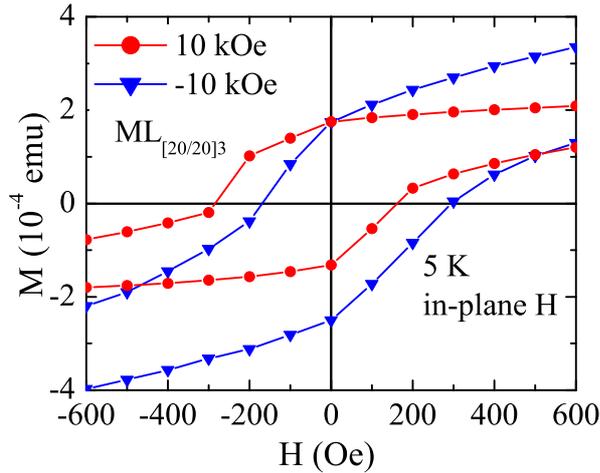}
	\caption{(color online) Expanded view of field dependent magnetization measured at 5 K following FC protocol with cooling field $H_{cool}$ = 10 kOe and -10 kOe for ML$_{[20/20]3}$ multilayer.}
	\label{fig:Fig4}
\end{figure}

We have calculated the exchange bias field (H$_{EB}$) and effective coercive field $H_c$ using $H_{EB}$ = ($H_c^L$ + $H_c^R$)/2 and $H_c$ = ($H_c^L$ - $H_c^R$)/2, respectively. For the $M(H)$ data collected at 5 K with field $\pm$ 10 kOe (Figure 3), we calculate $H_{EB}$ = 63 and $|H_c|$ = 178 Oe for ZFC and $H_{EB}$ = -64 and $|H_c|$ = 225 Oe for FC mode. We believe this EB effect in present multilayer is caused by an interface phenomena. For further confirmation, we have grown a single layer LSMO film ($\sim$ 50 nm) on STO and measured $M(H)$ at low temperature. The lower inset in Figure 3 shows close view FC $M(H)$ plot of LMSO film which presents a nearly symmetric $M(H)$ where we calculate $H_{EB}$ $\sim$ 10 Oe. Considering that $M(H)$ data are collected using a high field (14 T) superconducting magnet, this asymmetry can be regarded as an error limit of experimental accuracy or an insignificant effect. While EB effect is most commonly observed in FC condition where spins are biased to align at the point of magnetic transition, EB effect observed under both ZFC and FC condition in present system is quite interesting.

To further confirm the exchange bias effect in present multilayers, FC $M(H)$ loops have been recorded after cooling in both positive and negative magnetic field. For real EB effect, shifting of $M(H)$ loop will be in opposite direction of applied field. Figure 4 shows FC $M(H)$ data after cooling in +10 and -10 kOe field. As evident in figure, after cooling in +10 kOe the $M(H)$ loop is shifted toward negative field direction while for cooling in -10 kOe, it shifts to positive field direction. This opposite shifting confirms that observed EB in this multilayer is genuine. We have calculated $H_{EB}$ = 63 Oe and $|H_c|$ = 231 Oe for cooling field of -10 kOe, which closely match with the values ($|H_{EB}|$ = 64 Oe and $|H_c|$ = 225 Oe) for 10 kOe cooling field.

There have been only very few systems where the ZFC EB or spontaneous EB has been observed previously. The bifurcation between $M(T)$ in Figure 2 and shifting of $M(H)$ in Figure 3 suggest that there is an onset of AFM type interaction at interface through Mn-O-Ir exchange path. This, however, happens when SIO orders ferromagnetically at low temperature below $T_P$ as LSMO is already magnetically ordered from high temperature. To understand this, we have calculated $H_{EB}$ and $H_c$ as a function of temperature for both ZFC and FC protocol. Figure 5 shows $H_{EB}$ decreases very sharply with increasing temperature for both ZFC and FC data and at $T$ = 30 K, $H_{EB}$ almost vanishes. This temperature is close to $T_P$ in Figure 2. This fast decrease in $H_{EB}$ can be attributed to sharp weakening of interfacial exchange coupling between Mn and Ir ions. Th is also evident in the fact that $H_c^R$ does not change much but $H_c^L$ changes significantly with increasing temperature. This continuous evolution in asymmetry of $M(H)$ causes EB effect in present system.

\begin{figure}
	\centering
		\includegraphics[width=8cm]{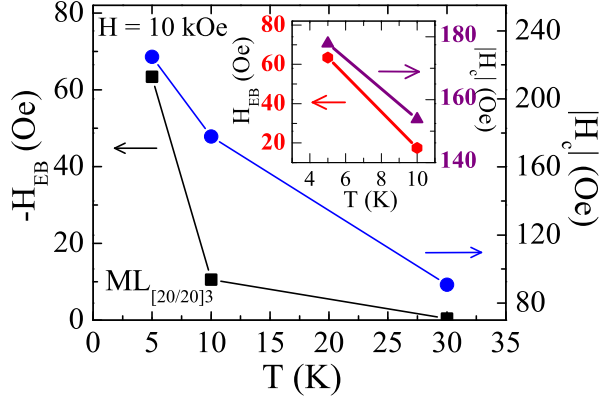}
	\caption{(color online) Exchange bias field (left axis) and coercive field (right axis) are shown as a function of temperature for ML$_{[20/20]3}$ multilayer for FC protocol. Inset shows values for ZFC protocol.}
	\label{fig:Fig5}
\end{figure}

To further understand the interrelation between present EB effect with interface, we have measured FC (1 T) $M(H)$ at 5 K for multilayer ML$_{[15/15]6}$ (Figure 6). We find $H_c^L$ = -576 and $H_c^R$ = 396 Oe which gives $H_{EB}$ = -90 and $|H_c|$ = 486 Oe. While though not proportionally, but these values are substantially higher than the respective values (-64 and 225 Oe) for ML$_{[20/20]3}$ multilayer which underlines the role of interface in present EB effect. It can be further noted that normalized moment (emu/cc) for ML$_{[15/15]6}$ multilayer is observed almost double compared to ML$_{[20/20]3}$ which is consistent with double number of layers in ML$_{[15/15]6}$ (not shown).

It is somehow believed that FM exchange coupling at interface gives negative EB effect while AFM type coupling leads to positive EB effect.\cite{nogues,nogues1} In this scenario, shifting of $M(H)$ loop in Figure 3 toward both negative and positive field axis in FC and ZFC measurements, respectively in same system is rather interesting. We consider that FM ordering of SIO film at low temperature initiates an AFM type Mn-O-Ir exchange coupling at interface. In general, nature of magnetic interaction in oxides depends on many factors such as, electronic structure of neighboring cations, bond angle and length, oxygen vacancies, etc. The lattice mismatch between LSMO and SIO is around 2\% which naturally induces strain at interface and can modify the bond angle and length at Ir-O-Mn exchange path. Similarly, electronic charge transfer across the interface, which is though limited to few atomic layers close to interface, can also influence the exchange coupling at interface. Recently, a charge transfer induced EB has been shown in FM/PM La$_{0.75}$Sr$_{0.25}$MnO$_3$/LaNiO$_3$ system which is rather unexpected.\cite{sanchez} However, such charge transfer would be an unlikely phenomena here as iridium occurs as Ir$^{4+}$ (5$d^5$) in SrIrO$_3$ which under the assumption of strong spin-orbit coupling effect will realize fully filled $J_{eff}$ = 3/2 quartet and partially filled $J_{eff}$ = 1/2 doublet state.\cite{kim1,kim2} This gives Ir$^{4+}$ a magnetically active $J_{eff}$ = 1/2 state. In this scenario, an interface charge transfer will give either Ir$^{3+}$ (5$d^6$) or Ir$^{5+}$ (5$d^4$) where both are supposed to be nonmagnetic as former will have fully filled $J_{eff}$ = 3/2 and 1/2 states and later will have only filled $J_{eff}$ = 3/2 state. Therefore, Ir$^{3+}$/Ir$^{5+}$ ion cannot support the magnetic interaction at interface, hence and EB effect.
\begin{figure}
	\centering
		\includegraphics[width=8cm]{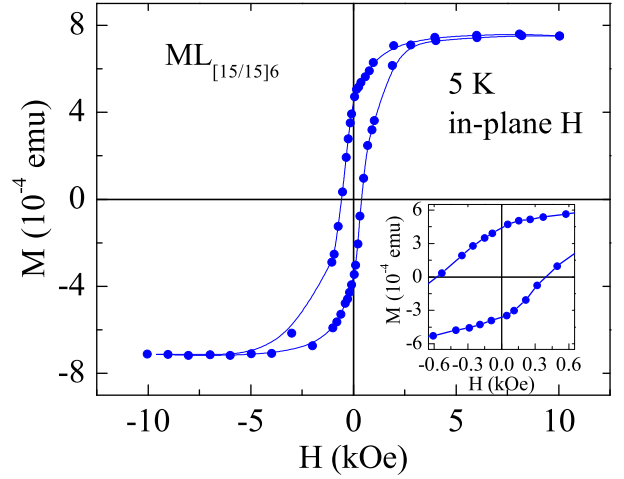}
	\caption{(color online) FC $M(H)$ plot ML$_{[15/15]6}$ multilayer at 5 K. inset shows expanded view close to origin.}
	\label{fig:Fig6}
\end{figure}

Nonetheless, individual layer of LSMO and SIO will have different magnetic anisotropy. Using x-ray magnetic circular dichroism (XMCD) measurements, a recent study has demonstrated high value of perpendicular magnetic anisotropy in LSMO/SIO superlattice which arises due to rotation of oxygen octahedra at interface.\cite{yi} In case of ZFC $M(H)$, although there is a naturally developed AFM type interface but the applied magnetic field during first application of magnetic field (virgin curve) aligns the AFM spins at interface. The effect will be prominent at interface as the interface will have relatively softer magnetic anisotropy. This induced FM interface would have high magnetic energy, hence, reversing the magnetic field would cost larger energy in rotating the FM spins. Therefore, a positive shifting of M(H) is realized with an AFM type interface coupling. On the other hand, during FC process the coupling of magnetic field with spins at interface would help the Zeeman interaction to win over the exchange coupling, therefore, interface would be FM like. In this case, cooling field is expected to have a serious ramification on the EB effect.

\begin{figure}
	\centering
		\includegraphics[width=8cm]{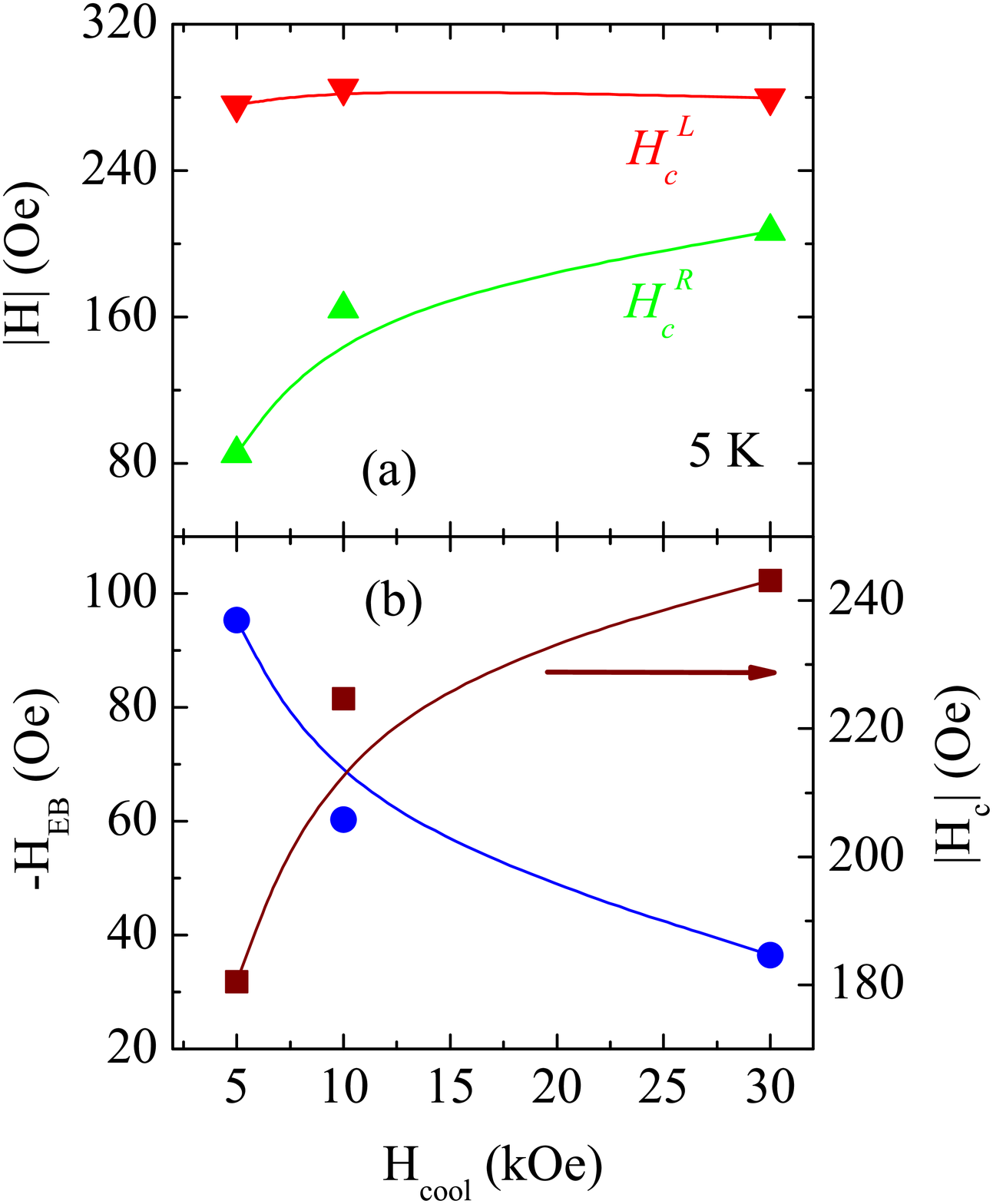}
	\caption{(color online) (a) shows variation of $H_c^L$ and $H_c^R$ with cooling field at 5 K for ML$_{[20/20]3}$ multilayer. (b) shows calculated $H_{EB}$ and $H_c$ as a function of cooling field. Lines are guide to eyes.}
	\label{fig:Fig7}
\end{figure}

To understand the effect of external field cooling on $H_{EB}$ and $H_c$, FC $M(H)$ measurements have been done at 5 K with different $H_{cool}$ (not shown). The saturation moment decreases continuously with decrease in $H_{cool}$, however, both the coercive fields respond differently. Figure 7a shows variation of both $H_c^L$ and $H_c^R$ for $H_{cool}$ =  5, 10 and 30 kOe. It is evident in figure that while $H_c^L$ remains almost constant, the value of $H_c^R$ increases continuously with $H_{cool}$. As $H_{cool}$ increases, the $H_c^R$ approaches $H_c^L$ yielding more symmetric $M(H)$ loop. Figure 7b shows variation of $H_{EB}$ and $H_c$ with $H_{cool}$. As expected, $H_{EB}$ decreases and $H_c$ increases monotonically with increasing $H_{cool}$. The negative EB in Figure 7 suggests an onset of FM unidirectional anisotropy at interface during field cooling. The higher magnetic fields destroy this unidirectional anisotropy as evident from increasing value of $H_c^R$ while $H_c^L$ remains constant.

\begin{figure}
	\centering
		\includegraphics[width=8cm]{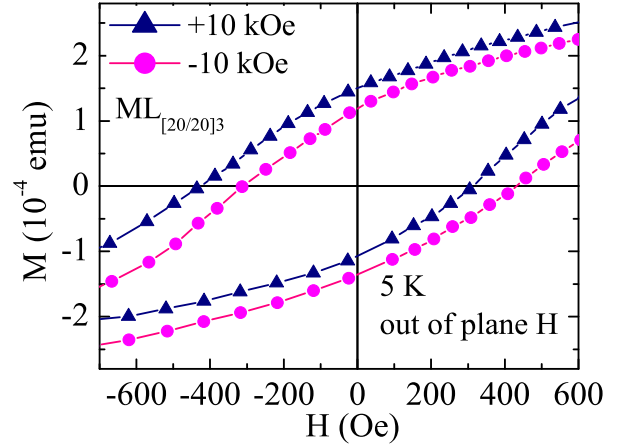}
	\caption{(color online) Expanded view of field dependent magnetization measured at 5 K following FC protocol with out-of-plane cooling field $H_{cool}$ = 10 kOe and -10 kOe for ML$_{[20/20]3}$ multilayer.}
	\label{fig:Fig8}
\end{figure}

To understand the out-of-plane magnetic asymmetry we have measured field cooled $M(H)$ for ML$_{[20/20]3}$ multilayer at 5 K. The field cooled $M(H)$ graph has been measured after cooling the multilayer in +10 kOe and -10 kOe field applied at 300 K, along  out-of-plane direction. The Fig. 8 shows $M(H)$ plot with $H_{cool}$ = 10 kOe and -10 kOe at 5 K. As evident in Fig. 8, the $M(H)$ loop for $H_{cool}$ = 10 kOe shifts towards negative field axis and the one with $H_{cool}$ = -10 kOe shifts towards positive field axis. This opposite shifting of $M(H)$ with cooling in positive and negative field, implies exchange bias effect. We have calculated $H_{EB}$ = -56 Oe for positive cooling field of 10 kOe, and $H_{EB}$ = 60 Oe for negative field cooling of -10 kOe. These values shows $H_{EB}$ almost similar to earlier obtained values, when magnetic field is in-plane direction.

As discussed, in addition to conventional FC negative EB, only few studies have reported ZFC positive and/or spontaneous EB effect, and in most of the cases it is observed in natural materials. Therefore, both FC and ZFC EB effect in present artificially made superlattices is quite noteworthy. Moreover, this multilayer exhibits higher $H_{EB}$ at lower $H_{cool}$ (Figure 7) where in other systems requirement of magnetic field is rather high. This study further provides a crucial information in terms of iridates that while bulk SrIrO$_3$ exhibits PM behavior throughout the temperature range, it develops magnetic ordering at low temperatures in thin films. The strain induced ferromagnetic ordering has been previously observed in $3d$ based CaRuO$_3$ film, where its bulk material shows PM and metallic behavior similiar to SrIrO$_3$.\cite{tripathi} This is important because while films and superlattics of 3$d$ and 4$d$ oxides have been extensively studied, the same for 5$d$ materials is less explored, and this information will be helpful to understand these class of materials.

\section{Conclusion}
In conclusion, epitaxial multilayer of 3$d$-5$d$ based oxide [LSMO/SIO]$_3$ is grown on STO (100) substrate. We observe both FC-negative and ZFC spontaneous EB effect in this rather unconventional FM/PM multilayer. The EB effect decreases with increasing temperature and completely disappears when it reaches to interface magnetic ordering temperature ($\sim$ 40 K). Moreover, EB field decreases with increasing cooling field in FC process. The ZFC EB effect originates an AFM type Mn-O-Ir exchange interaction at interface which is favored by FM ordering of SIO film. The applied field during FC helps the Zeeman interaction to win over the exchange coupling which induces FM interface and FC negative EB effect. This is important that present unconventional EB behavior probes the magnetic nature of SIO at low temperature in form of thin film.

\section{Acknowledgment}   
We thank SQUID facility at IIT Delhi and Ratnamala Chatterjee for the magnetic measurements. We are thankful to AIRF, JNU for XRD and magnetization measurements. We acknowledge SERB, DST for funding Excimer pulse laser, UPE-II, UGC for funding deposition chamber and DST-PURSE for financial support.

\end{document}